\documentclass[english,aps, prb, twocolumn, superscriptaddress,showpacs]{revtex4-1}
\usepackage{lmodern}
\usepackage[T1]{fontenc}
\usepackage[latin9]{inputenc}
\usepackage{graphicx}
\usepackage{natbib,color}

\makeatletter

\@ifundefined{textcolor}{}
{%
 \definecolor{BLACK}{gray}{0}
 \definecolor{WHITE}{gray}{1}
 \definecolor{RED}{rgb}{1,0,0}
 \definecolor{GREEN}{rgb}{0,1,0}
 \definecolor{BLUE}{rgb}{0,0,1}
 \definecolor{CYAN}{cmyk}{1,0,0,0}
 \definecolor{MAGENTA}{cmyk}{0,1,0,0}
 \definecolor{YELLOW}{cmyk}{0,0,1,0}
 }

\makeatother

\usepackage{babel}

\begin{document}

\title{Spin fluctuations in LiFeAs observed by neutron scattering}

\author{A. E. Taylor }

\affiliation{Department of Physics, University of Oxford, Clarendon Laboratory,
Parks Road, Oxford, OX1 3PU, United Kingdom}

\author{M. J. Pitcher}

\affiliation{Inorganic Chemistry Laboratory, University of Oxford, South Parks
Road, Oxford OX1 3QR, United Kingdom}

\author{R. A. Ewings}

\author{T. G. Perring}

\affiliation{ISIS Facility, Rutherford Appleton Laboratory, STFC, Chilton, Didcot,
Oxon, OX11 0QX, United Kingdom}

\author{S. J. Clarke}

\affiliation{Department of Chemistry, University of Oxford, Inorganic Chemistry Laboratory, South Parks
Road, Oxford OX1 3QR, United Kingdom}

\author{A. T. Boothroyd}

\affiliation{Department of Physics, University of Oxford, Clarendon Laboratory,
Parks Road, Oxford, OX1 3PU, United Kingdom}

\begin{abstract}
We report neutron inelastic scattering measurements on the stoichiometric iron-based superconductor LiFeAs.  We find evidence for (i) magnetic scattering consistent with strong antiferromagnetic fluctuations, and (ii) an increase in intensity in the superconducting state at low energies, similar to the resonant magnetic excitation observed in other iron-based superconductors. The results do not support a recent theoretical prediction of spin-triplet $p$-wave superconductivity in LiFeAs, and instead suggest that the mechanism of superconductivity is similar to that in the other iron-based superconductors.
\end{abstract}
\maketitle

LiFeAs exhibits several properties which apparently set it apart from the rest of the iron-based superconductors. Unlike other stoichiometric iron arsenide compounds, LiFeAs is an intrinsic superconductor with relatively high transition temperature $T_{\mathrm{c}} \approx 17$~K without the need for carrier doping or application of pressure to induce superconductivity\cite{LiFeAs_Guloy2008,LiFeAs_Clarke2008,LiFeAs_Jin2008,LiFeAs+NaFeAs_Xue2009}. Accordingly, and again in contrast to other iron-based superconductors, stoichiometric LiFeAs does not undergo a tetragonal-to-orthorhombic phase transition, even up to pressures of 20~GPa (Refs.~\onlinecite{LiFeAs_Pressure_Clarke2009} and \onlinecite{LiFeAs_Jin2009}), and does not exhibit
spin density wave (SDW) order close to, or coexisting with, the superconducting
regime\cite{LiFeAs+NaFeAs_Xue2009,LiFeAs_Guloy2008,LiFeAs_muSR_Clarke2009}. This anomalous behavior raises the possibility that some aspects of the superconductivity in LiFeAs might be different from the other iron-based superconductors.

The lack of SDW order in LiFeAs has recently been discussed in relation to Fermi surface data obtained by angle-resolved photoemission spectroscopy (ARPES)\cite{LiFeAs_Buechner2010}. SDW order in the iron-based superconductors is generally attributed to strong Fermi surface nesting between electron and hole pockets separated by the antiferromagnetic wavevector ${\bf Q}_{\rm AF}$ (see Ref.~\onlinecite{Singh_review2009}). In the superconducting compounds, SDW order is suppressed or completely absent, but antiferromagnetic fluctuations persist into the superconducting state and have been observed by inelastic neutron scattering\cite{Magnetism_review2010}. A prominent feature of the spin fluctuation spectrum is a magnetic resonance which appears at temperatures below $T_{\mathrm{c}}$ and at the wavevector ${\bf Q}_{\rm AF}$. This feature, first observed in (Ba,K)Fe$_2$As$_2$ (Ref.~\onlinecite{(BaK)Fe2As2_Guidi2008}) and subsequently in many other Fe-based superconductors\cite{Magnetism_review2010}, is consistent with a dominant spin-singlet $s^{\pm}$ pairing symmetry\cite{Theory_PRL_S+-_Du2008,Theory_PRB_Scalapino2008,Theory_PRB_Eremin2008} as also suggested by many other experiments\cite{Johnston_review2010}.

The Fermi surface of LiFeAs, however, is found to display poor nesting properties\cite{LiFeAs_Buechner2010}. This suggests that differences in its physical properties compared with other Fe-based superconductors could be due to differences in the electronic structure, resulting at least partly from the short Fe--Fe distance compared with most other Fe-based superconductors~\cite{LiFeAs_CompControl_Clarke2010}. An electronic structure model for LiFeAs based on the ARPES results supports this notion\cite{LiFeAs_TripletTheo_Brink2011}.  The model predicts that SDW order is absent, and instead finds almost ferromagnetic fluctuations which drive an instability towards spin-triplet $p$-wave superconductivity. The character of the spin fluctuations is therefore pivotal to
the superconducting pairing state, according to these models, and so experimental measurements of the spin fluctuation spectrum of LiFeAs are of great interest.

Here we report a neutron inelastic scattering study of the momentum-resolved magnetic spectrum of polycrystalline LiFeAs. The data provide clear evidence for strong magnetic fluctuations with a characteristic wavevector coincident with (or close to) ${\bf Q}_{\rm AF}$, consistent with NMR data\cite{LiFeAs_NMR_PRB_Arcon2010,LiFeAs_NMR_PRB_Weiqiang2010}. We also observe an increase in intensity within a range of energies around $E \approx 8$~meV on cooling below $T_{\rm c}$, consistent with a superconductivity-induced spin resonance peak.

\begin{figure}

\includegraphics[width=0.9\columnwidth]{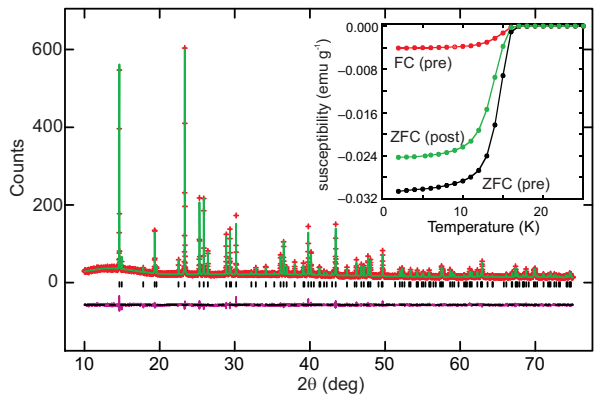}
\caption{(Color online) Rietveld refinement of x-ray powder diffraction data collected on instrument I11 (Diamond Light Source) showing that the sample is phase pure with lattice parameters $a = 3.77657(1)\,\mathrm{\AA}$, and $c = 6.35527(2)\,\mathrm{\AA}$ corresponding to stoichiometric LiFeAs (Ref.~\onlinecite{LiFeAs_CompControl_Clarke2010}).
The inset shows the magnetic susceptibility of portions of the sample measured in an applied field of $50$\, Oe prior to the neutron scattering measurements (``pre'') under zero-field-cooled (ZFC) and field-cooled (FC) conditions, and after the measurements (``post''; ZFC measurement only). The values of the susceptibilities correspond to bulk diamagnetism (see Ref.~\onlinecite{LiFeAs_Clarke2008}).
\label{fig:XRD}}

\end{figure}

Polycrystalline LiFeAs was prepared from high-purity elemental reagents
(>99.9\%) by the methods reported elsewhere~\cite{LiFeAs_CompControl_Clarke2010}. The sample
``MP127'' described in Ref.~\onlinecite{LiFeAs_CompControl_Clarke2010} was used for this experiment. Joint
synchrotron x-ray powder diffraction (see Fig.~\ref{fig:XRD}) and neutron powder diffraction
refinements (space group $P4/nmm$, lattice parameters $a=3.777\,\mathrm{\AA}$
and $c=6.356\,\mathrm{\AA}$) established that the material was phase pure and that there was no detectable substitution of Li by Fe which has been shown to destroy superconductivity even at the 2\% level\cite{LiFeAs_CompControl_Clarke2010}. Magnetic susceptibility measurements
made by SQUID magnetometry confirmed that the sample is a bulk superconductor
with a sharp onset of superconductivity at $T_{\mathrm{c}}=17\,\mathrm{K}$ (Fig.~\ref{fig:XRD}). 

The inelastic neutron scattering experiments were performed on the MERLIN chopper spectrometer at the
ISIS Facility\cite{MERLIN_Coleman2006}. Approximately $7.5\,\mathrm{g}$
of LiFeAs powder was sealed inside a cylindrical aluminium can and
mounted in a top-loading closed-cycle refrigerator. Due to the extreme air sensitivity of the sample, all handling was done in an inert atmosphere. After the experiment the sample was re-checked by x-ray diffraction and magnetometry and its properties were found to be the same as before the experiment.  Spectra were recorded 
with incident neutron energies $E_{\rm i}=15$, 25 and 50~meV, and
at a number of temperatures between 6\,K and 34\,K. A short run was also performed at room temperature. The scattering from a standard
vanadium sample was used to normalize the spectra and to place them
on an absolute intensity scale with units $\mathrm{mb}\,\mathrm{sr}^{-1}\,\mathrm{meV^{-1}}\,\mathrm{f.u.}^{-1}$,
where $1\,\mathrm{mb}=10^{-31}\,\mathrm{m}^{2}$ and f.u. stands for
``formula unit'' (of LiFeAs).

\begin{figure}
\includegraphics[clip,width=0.9\columnwidth]{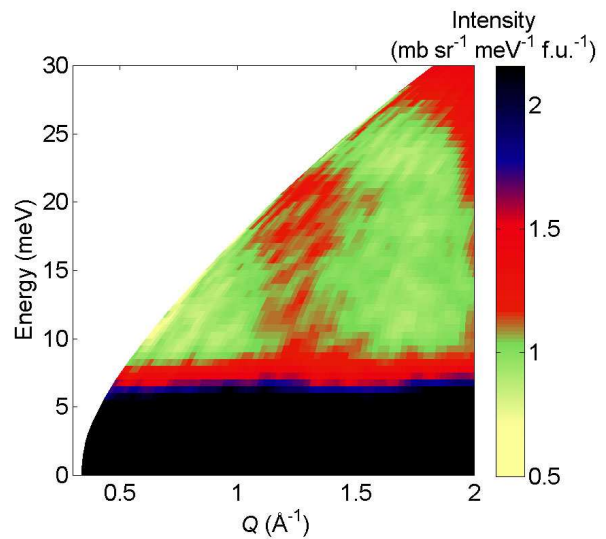}

\caption{\label{fig:50meV_6K_Slice}(Color online) Neutron scattering spectrum of polycrystalline LiFeAs.
The data were recorded on MERLIN at a temperature of $6\,$K with an incident neutron
energy of $50\,$meV.}

\end{figure}

The general features of the data are illustrated in Fig.~\ref{fig:50meV_6K_Slice},
which is a color plot of the measured inelastic scattering intensity as a function
of momentum transfer, $Q$, and energy transfer, $E$. At low energies there is strong scattering from the elastic line (coherent and incoherent scattering) and from phonons. However, a steep column of
scattering centered on $Q \approx 1.2\,\mathrm{\AA}{}^{-1}$ stands out
from the phonon background and extends in energy throughout the accessible
region of $(Q, E)$ space. The spectrum bears a very close resemblance to that of polycrystalline BaFe$_2$As$_2$ (Ref.~\onlinecite{BaFe2As2_Boothroyd2008}) and (Ba,K)Fe$_2$As$_2$ (Ref.~\onlinecite{(BaK)Fe2As2_Guidi2008}), and based on this we can confidently attribute the column of scattering at $Q \approx 1.2\,\mathrm{\AA}{}^{-1}$ to magnetic fluctuations.

\begin{figure}
\includegraphics[width=0.9\columnwidth]{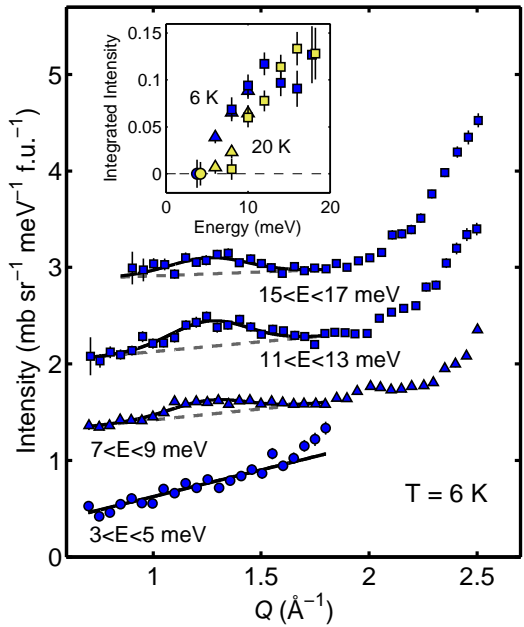}

\caption{(Color online) Constant-energy cuts showing the magnetic signal at $Q \approx 1.2\,\AA^{-1}$ at different energies for a fixed temperature of 6\,K (main panel). Data were averaged over the indicated energy intervals. Successive cuts are displaced vertically for clarity. The symbols represent different neutron incident energies: squares $E_{\rm i}=50\,\mathrm{meV}$, triangles $E_{\rm i}=25\,\mathrm{meV}$, circles $E_{\rm i}=15\,\mathrm{meV}$. The full lines are fits are to Gaussian peaks centered at $Q=1.24\,\AA^{-1}$ with standard deviation $0.18\,\AA^{-1}$ on a linear background (broken lines). The inset shows the integrated intensities (in $\mathrm{mb}\,\mathrm{sr}^{-1}\,\mathrm{meV^{-1}}\,\mathrm{f.u.}^{-1}\,\AA^{-1})$ of the fitted Gaussian peaks as a function of energy for temperatures of 6\,K and 20\,K. The 6\,K and 20\,K points at 4\,meV and 18\,meV are in reality almost coincident, but have been separated horizontally by 0.5\,meV to make them visible. The elastic energy resolution (full-width at half-maximum) for the three incident energies is 4.4\,meV (50\,meV), 2.1\,meV (25\,meV) and 1.1\,meV (15\,meV), }\label{fig:IIvsE_cuts}

\end{figure}

Figure~\ref{fig:IIvsE_cuts} presents a selection of cuts taken through the data at different energies. The cuts all contain a rising signal at higher $Q$ due to phonon scattering and (apart from at the lowest energy) a peak centered on $Q \approx 1.2\,\mathrm{\AA}^{-1}$.
To analyze the peak quantitatively we fitted the cuts to a Gaussian function on a linear background. Initial fits were made in which the width, center and amplitude of the Gaussian, and the slope and intercept of the linear background, were allowed to vary. The peak center and width were found not to vary significantly with energy or temperature, with average values $Q=1.24\pm0.02\,\mathrm{\AA}^{-1}$ and $\sigma=0.18\pm0.02\,\mathrm{{\AA}}^{-1}$ (standard deviation), respectively. For all subsequent fits
we fixed the center and width of the Gaussian to these values. This is physically reasonable because the magnetic interactions are strong and the dispersion very steep --- see Fig~\ref{fig:50meV_6K_Slice} --- as found in other Fe-based superconductors\cite{Magnetism_review2010}. In most cases these constraints do not significantly affect the values of the fitted intensities, but at the lowest energies where the signal is small, and at the highest energies where there is limited data on the background on the low $Q$ side of the peak, they reduce the uncertainties in the fitted peak intensities. 

%

The inset of Fig.~\ref{fig:IIvsE_cuts} is a plot of the integrated intensities of the fitted peaks as a function of energy for the accessible range of energies. Data are shown for temperatures of 6\,K ($T<T_{\rm c}$) and 20\,K ($T>T_{\rm c}$). At 20\,K there is no measurable intensity up to an energy of about 5\,meV, at which point the intensity increases sharply with energy. This behavior indicates that there is a spin gap in the normal state of about 10\,meV. At 6\,K there is additional intensity in the energy range between about 4\,meV and 12 meV.

We also performed fits with a powder-averaged two-dimensional Gaussian function, a model that assumes no variation in intensity with out-of-plane momentum or with energy. This method put the center
of the Gaussian at $Q=1.14\pm0.02\,\mathrm{\AA}^{-1}$. In reality we
expect the magnetic correlations to be quasi-two-dimensional (intermediate between two- and three-dimensional due to non-negligible $c$-axis coupling, as observed in other iron arsenides), so we expect
the true characteristic wavevector of the fluctuations to be in the range $Q=1.14-1.24\,\mathrm{\AA}^{-1}$,
consistent with the antiferromagnetic wavevector ${\bf Q}_{\rm AF} = (0.5,0.5,0)$ r.l.u., for which  $Q_{\mathrm{AF}}=1.18\,\mathrm{\AA}^{-1}$.

\begin{figure}

\includegraphics[width=0.9\columnwidth]{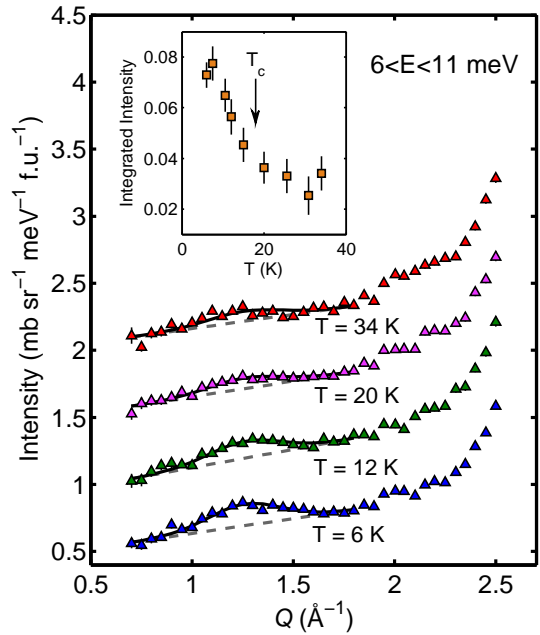}
\caption{(Color online) Constant-energy cuts showing the temperature dependence of the magnetic signal at $Q \approx 1.2\,\AA^{-1}$. Data in each cut were recorded with an incident energy $E_{\rm i}=25\,$meV and have been averaged over an energy range 6--11\,meV. Successive cuts are displaced vertically for clarity. Full lines are fits are to Gaussian peaks centered at $Q=1.24\,\AA^{-1}$ on a linear background (broken lines). The inset shows the integrated intensity of the fitted Gaussian peaks as a function of temperature.
\label{fig:IIvsT_cuts}}

\end{figure}

Figure \ref{fig:IIvsT_cuts} shows a series of cuts taken through data collected at different temperatures, with neutrons of incident energy 25\,meV. All of these cuts are averaged over an energy interval of 6--11\,meV. Small peaks visible near $Q \approx 2\,\mathrm{\AA}^{-1}$ are from phonons since they increase with temperature. We used the same fitting procedure as described above to obtain the temperature variation of the magnetic signal. Again, the center and width of the Gaussian were fixed, and only the peak amplitude and linear background were varied. The inset of Fig.~\ref{fig:IIvsT_cuts} shows the integrated intensity of the fitted Gaussian peak as a function of temperature. The integrated intensity is seen to increase sharply below the superconducting transition temperature $T_{\mathrm{c}}=17\,\mathrm{K}$.

The measurements reported here have revealed two pieces of new information about LiFeAs. The first is that cooperative spin fluctuations exist and have a characteristic wavevector close to $Q_{\mathrm{AF}}=1.18\,\mathrm{\AA}^{-1}$. While we cannot rule out a small incommensurate splitting of order $0.1 {\AA}^{-1}$, the data indicate that LiFeAs has similar spatial magnetic correlations to those in the other iron arsenide superconductors. The absolute values of the integrated intensity shown in the inset of Fig.~\ref{fig:IIvsE_cuts} are comparable to those obtained for BaFe$_{2}$As$_{2}$ by the same technique\cite{BaFe2As2_Boothroyd2008},
which suggests that the size of the fluctuating moments are similar in the two materials.

At the same time, we find no evidence for near-ferromagnetic fluctuations. The closest ferromagnetic wavevectors to the observed magnetic signal are $Q_{\mathrm{(001)}} = 0.99\,\mathrm{\AA}^{-1}$ and $Q_{\mathrm{(002)}}=1.98\,\mathrm{\AA}^{-1}$, and since these are Fourier components of the iron primitive lattice we would expect ferromagnetic fluctuations to give strong magnetic scattering at these wavevectors. It is evident from Figs.~\ref{fig:50meV_6K_Slice}--\ref{fig:IIvsT_cuts} that no detectable signals exist at either of these positions. The direct observation of SDW fluctuations, as opposed to ferromagnetic fluctuations, in the momentum-resolved magnetic spectrum is consistent with the interpretation of NMR data on LiFeAs\cite{LiFeAs_NMR_PRB_Arcon2010,LiFeAs_NMR_PRB_Weiqiang2010}.


The existence of strong SDW fluctuations, and lack of ferromagnetic fluctuations, is perhaps surprising given (i) the results from ARPES which indicate poor nesting between electron and hole pockets separated by ${\bf Q}_{\rm AF}$, Ref.~\onlinecite{LiFeAs_Buechner2010}, and (ii) the theoretical calculations based on the ARPES results which predict near-ferromagnetic fluctuations to be the dominant pairing interaction\cite{LiFeAs_TripletTheo_Brink2011}. On the other hand, more detailed calculations using a functional renormalization group method indicate that although ferromagnetic fluctuations are present as a competing instability, antiferromagnetic fluctuations dominate at the low energy scales relevant for superconductivity\cite{LiFeAs_DFT_AFM_Hanke2011} and could drive an instability towards $s^{\pm}$ superconducting pairing in LiFeAs, similar to that generally thought to occur in other iron-based superconductors.

The second notable feature of our results is the increase in spectral weight on cooling below $T\approx T_{\mathrm{c}}$ for energies in the vicinity of the spin gap, as shown in the insets to Figs.~\ref{fig:IIvsE_cuts} and ~\ref{fig:IIvsT_cuts}.  This behavior is qualitatively consistent with a superconductivity-induced magnetic resonance, as reported in the 122 and 11 iron-based superconductors\cite{Magnetism_review2010}, and most recently in the 1111 family\cite{LaFeAsO_Shamoto2010}. An approximate scaling has been found between the resonance energy $E_{\rm r}$ and $T_{\mathrm{c}}$, such that $E_{\rm r}/k_{\rm B}T_{\mathrm{c}} = 4.5-5.5$ (Ref.~\onlinecite{Magnetism_review2010}).  If this scaling is applied to LiFeAs ($T_{\mathrm{c}} = 17$\,K) then it predicts a magnetic resonance near 7.5\,meV, which is in the middle of the range in which additional intensity is observed below $T\approx T_{\mathrm{c}}$.  On balance, therefore, the results point towards the existence of a magnetic resonance below $T_{\mathrm{c}}$ in LiFeAs.


In conclusion, we have obtained two key results which provide firm evidence that LiFeAs behaves in a similar way to the other iron pnictides and is not in fact anomalous. We have observed scattering from strong spin fluctuations at or close to the antiferromagnetic wavevector, and we find evidence for the existence of a superconductivity-induced magnetic resonance at around 8\,meV. These results suggest that the mechanism of superconductivity in LiFeAs is similar to that in other iron-based superconductors.

This work was supported by the Engineering \& Physical Sciences Research Council and the Science \& Technology Facilities Council.

\bibliographystyle{apsrev4-1}
\bibliography{LiFeAsPaperBib}
\end{document}